\documentclass[lettersize,journal]{IEEEtran}
\usepackage{amsmath,amsfonts,amssymb}
\usepackage{algorithm}
\usepackage{algpseudocode}
\usepackage{array}
\usepackage[caption=false,font=normalsize,labelfont=sf,textfont=sf]{subfig}
\usepackage{textcomp}
\usepackage{stfloats}
\usepackage{url}
\usepackage{verbatim}
\usepackage{graphicx}
\usepackage{multirow}
\usepackage{cite}
\usepackage[font=small]{caption}
\usepackage{xcolor}
\usepackage{hyperref}
\usepackage{dsfont}
\usepackage{mathtools}
\usepackage{booktabs}
\begin{document}

\title{Underground Power Distribution System Restoration\\ Using Inverter Based Resources}

\author{IEEE Publication Technology,~\IEEEmembership{Staff,~IEEE,}
}

\author{Wenlong~Shi,~\IEEEmembership{Member,~IEEE},~Hongyi~Li,~\IEEEmembership{Member,~IEEE},~and~Zhaoyu~Wang,~\IEEEmembership{Senior  Member,~IEEE}       
	\thanks{This work was partially supported by the Power System Engineering and Research Center under Grant PSERC S-110, the U.S. Department of Energy Office of Energy Efficiency and Renewable Energy under Grant DE-EE0011234, and the National Science Foundation under Grant ECCS 2042314.}
	\thanks{The authors are with the Department of Electrical and Computer Engineering,
		Iowa State University, Ames, IA 50011 USA (e-mail:
		wshi5@iastate.edu; hongyili@iastate.edu; wzy@iastate.edu).}
\thanks{(Corresponding author: Zhaoyu Wang.)}
}



\setlength{\abovedisplayskip}{3.5pt}
\setlength{\belowdisplayskip}{3.5pt}

\maketitle

\begin{abstract}
Underground power distribution systems (PDSs) are increasingly deployed in urban areas. The integration of smart devices including smart switchgears, pad-mounted distribution transformers and inverter‐based resources (IBRs) enhance system resilience, however simultaneously introducing unique challenges. The challenges include inrush currents caused by trapped charges in underground cables, ferroresonance in distribution transformers during energization, and three-phase load imbalance resulting from single-phase underground laterals. To address these issues, this paper proposes an underground PDS restoration framework using IBRs. Firstly, an underground cable energization model is developed to quantify  inrush current by analyzing voltage differences across both  switchgear terminals. Secondly, a distribution transformer energization model is proposed to evaluate ferroresonance using Q-factor constraints based on underground cable  capacitance and damping resistance.  Thirdly, a phase-swapping model is proposed to improve load balancing by dynamically reassigning lateral-phase connections through smart switchgears. The proposed models are further integrated into a mixed-integer nonlinear programming (MINLP) formulation to maximize the total weighted restored load while constraining inrush currents, ferroresonance, and phase imbalance. To address the nonlinearity induced by impedance matrix reordering during phase swapping, a permutation-based linearization technique is proposed. Finally, case studies on an underground PDS established based on IEEE 123-Node Test Feeder validate the effectiveness of the proposed strategy in improving uderground PDS restoration performance.

\end{abstract}


\section*{Nomenclature}

\noindent\textit{A. Parameters}

\begin{description}[\IEEEsetlabelwidth{$(\cdot)_r + \textbf{i}(\cdot)_i$}]
	\item[$C_g^\text{eq}$] The equivalent capacitance of underground cables connected to switchgear $g$.
	\item[$f^{\text{nom}}$] Nominal frequency, 60Hz.
	\item[$L_g^\text{osc}$] The equivalent inductance of transformers that meets ferroresonance condition.
	\item[$\mathcal{Q}_g^\text{max}$] Maximum Q-factor for safe energization.
	\item[$R_g^{t,\text{dp}}$] Damping resistance downstream of switchgear $g$ at time $t$.
	\item[$\overline{\boldsymbol{P}}_{j}^{t,\text{RES}}$] Three-phase forecasting active power output of renewable energy sources at node $j$ at time $t$.
	\item[$\mathcal{E}_k^\text{max}$] Maximum SOC of ESS at node $k$.
    \item[$\eta_k^{ch},\eta_k^{dis}$] Charging and discharging efficiencies.

\end{description}

\noindent\textit{B. Variables}

\begin{description}[\IEEEsetlabelwidth{$(\cdot)_r + \textbf{i}(\cdot)_i$}]
	\item[$c_k^t,d_k^t$] Binary variables, 1 meaning that ESS $k$ is being charged or discharged at time $t$.
	\item[$\boldsymbol{C}_{k}^{t},\boldsymbol{D}_{k}^{t}$] Three-phase charging and discharging power of ESS $k$ at time $t$.
	\item[$f_{ij}$] Fictitious commodity flow on line  $(i,j)$.
	\item[$\tilde{\boldsymbol{I}}_{ijk}^{t}$] Three-phase squared current magnitude on line $(i,j)$ with respect to ESS $k$ at time $t$.
	\item[$\boldsymbol{p}_{j}^{t},\boldsymbol{q}_{j}^{t}$] Three-phase active and reactive load demand of node $j$ at time $t$.
	\item[$\boldsymbol{P}_{j}^{t,\text{RES}}$]  Active power injection of solar PV and wind turbine of node $j$ at time $t$.
	\item[$\boldsymbol{Q}_{k}^{t}$] Three-phase reactive power support of ESS $k$.
	\item[$\boldsymbol{Q}_{j}^{t,\text{RES}}$] Solar PV and wind turbine reactive power support of node $j$ at time $t$.
	\item[$\boldsymbol{\mathcal{S}}^t_{g}$] Phase swapping matrix of switchgear $g$ at $t$.
	\item[$u_{ik}$] Binary variable, 1 if node $i$ is covered by the microgrid established by ESS $k$.
	\item[$\tilde{\boldsymbol{V}}_{ik}^{t}$] Three-phase squared voltage magnitude of node $i$ with respect to ESS  $k$ at time $t$.
	    
    \item[$\alpha_g^t$] Binary variable, 1  meaning transformers downstream switchgear of $g$ can be energized at $t$.
    \item[$\gamma_{ij}$] Binary variable, 0 if switch $(i,j)$ is opened.
    \item[$\Gamma_{ij}$] Binary variable, 0 if line $(i,j)$ in the ficticious graph is disconnected to maintain radiality.
    \item[$\beta_{g}^t$] Binary variable, 0 if switchgear $g$ is opened.
    \item[$\mathcal{E}_{k}^{t}$] State of Charge of ESS $k$  at time $t$.    
    \item[$\mathcal{Q}_g^t$] Q-factor downstream switchgear $g$ at time $t$.
    \item[$\Delta \boldsymbol{\mathcal{S}}^t_{g}$] Auxiliary matrix, representing the difference of phase swapping matrix between $t$ and $t-1$.
    \item[$\Delta {V}_{ijk,\varphi}^t$] The voltage difference of switchgear $g$ between nodes $i$ and $j$ in terms of phase $\varphi$.
    \item[$\Delta \theta_{ij,\varphi}^t$] The phase $\varphi$ angle difference of the switchgear between nodes $i$ and $j$.

\end{description}

\noindent\textit{C. Indices and Sets}

\begin{description}[\IEEEsetlabelwidth{$(\cdot)_r + \textbf{i}(\cdot)_i$}]
    \item[$v\in\mathds{V}$] Variations of line impedance reordering.
    \item[$\varphi,\phi\in\Phi$] Index of phase in three-phase systems.
    \item[$\mathds{C}_{ik}$] Set of child nodes of $i$ with respect to ESS $k$.
    \item[$\mathds{K}$] Set of ESS nodes in the network.
    \item[$\mathds{L}_g$] Set of lines downstream switchgear $g$.
    \item[$\mathds{N}_g$] Set of nodes downstream switchgear $g$.
    \item[$\mathds{Z}$] Sets of solar PV and wind turbine nodes.

\end{description}

\section{Introduction}

\IEEEPARstart{U}{nderground} power distribution systems (PDSs) are widely adopted in urban areas. From an aesthetic perspective, undergrounding power lines can eliminate the visual clutter of overhead lines \cite{dugstad2023place}. While for professionals in power systems from both industry and academia, underground PDSs can reduce the failure rate and enhance the system's reliability and resilience, especially against extreme weather events \cite{wang2025enhancing}. Due to the fact that vulnerability is the most critical consideration that is addressed by underground PDSs, their performance and utilization require scientific research.

In literature, research on underground PDSs mainly focuses on risk assessment and hardening strategies. Specifically, risk assessment evaluates the physical  performance of underground components, such as cable insulation, mechanical damage, and thermal stress, and uses fragility curves to quantify the failure probability \cite{10843867,poudel2019probabilistic,kongar2017seismic}. Moreover, hardening strategies investigates how to determine the optimal underground solution to improve system robustness. Broadly speaking, it can be categorized as empirical, stochastic and heuristic approaches. Among these, empirical approaches compare various hardening decisions via iterative simulation and assessment \cite{ahmad2023towards}. Stochastic approaches employ optimization frameworks to account for  damage uncertainties \cite{zhang2020multi,wang2022two,gan2021tri,hou2023resilience,chen2023two,li2023distributionally}. Heuristic approaches can improve computational efficiency, while maintaining solution accuracy \cite{pavon2019optimal,valenzuela2019planning}. Nonetheless, these methods are developed for long-term planning, the potentials and challenges of underground PDSs in terms of load restoration remain underexplored.

The potentials and challenges of underground PDSs represent the two different sides of the same coin. On the potential side, underground PDSs benefit from the grid evolution toward intelligence, and decentralization. For example, pad-mounted smart switchgears facilitate smarter operations in underground PDSs. The switches, equipped with sensors and microprocessors, not only perform opening and closing functions, but are also capable of measuring the voltage phasor at the two ends \cite{10858306}. The phase swapping of smart switchgears, namely reconnecting a single-phase underground lateral to another phase on the main feeder, can further enhance the restoration \cite{lin2005heuristic,grigoracs2023phase}. Another potential stems from the integration of inverter based resources (IBRs), such as energy storage systems (ESSs), solar photovoltaics (PVs) and wind turbines (WTs),  which enables a bottom-up blackstart restoration \cite{liu2023utilizing}. During this process, the grid-forming inverters regulate voltage and frequency without the need of traditional synchronous generators (SG). Also, the grid-following inverters  offer an additional power support, in coordination with grid-forming inverters.

However, as these smart devices are developed and deployed within underground PDSs, their utilization in terms of restoration introduces unique challenges:

\begin{enumerate}
	\item Underground cables typically have much higher capacitance compared with overhead lines, resulting in trapped charge. It can persist for a long time following an outage, causing inrush current when energizing the underground cables through switchgears \cite{pordanjani2016discharge}. Although the duration of this high inrush is typically very short, its effects are non-negligible. Without effective control, it can impose mechanical, thermal, and dielectric stresses on electrical devices, which will be degraded or failed over time.

	\item The capacitance of underground cables and the nonlinear magnetizing inductance of pad-mounted distribution transformers collectively form a resonant LC circuit on the primary side. During the energization, a phenomenon called ferroresonance can be induced \cite{pordanjani2021single}. It can result in severe voltage spikes, with transformer growling and even transformer failures. This ferroresonance should be managed carefully to achieve a successful restoration.

	\item Underground PDSs are typically three-phase unbalanced due to the widespread use of single-phase underground laterals. This feature imposes  stress on IBRs, which depend on control strategies to maintain symmetrical phase outputs without mechanical inertia \cite{sochor2017theoretical}. In other words, IBRs are less tolerant to load imbalances among phases, compared with traditional synchronous generators. Even though some IBRs are designed to deal with negative-sequence components, their capabilities are very limited.

\end{enumerate}

Based on the above discussions, we observe that the underground PDS restoration using IBRs needs further investigation. In particular, in terms of new challenges such as inrush current, ferroresonance and load imbalance, the optimal solutions must be determined to ensure the safe operation of smart devices. To this end, this paper proposes an underground PDS restoration strategy. The main contributions are threefolds:

\begin{enumerate}
	\item Corresponding to the three unique challenges in underground PDSs, three models are proposed. The first one is underground cable energization model, where the inrush current is captured based on the voltage difference across two terminals of switches. The second one is distribution transformer energization model, where ferroresonance is evaluated via damping levels and Q-factor. The third one is phase swapping model for smart switchgear operation, which reconnects underground laterals to the phases on main feeders for load balancing. 

	\item A mixed-integer nonlinear programming (MINLP) formulation is proposed for the underground PDS restoration problem. It aims to determine the optimal timing for switchgear reconnection and transformer energization to maximize the multi-period total weighted restored load, while imposing operational constraints on inrush current, ferroresonance, and load imbalance.
	
	\item To address the nonlinearity introduced by the process of smart switchgear reconnection, a  linearization technique is developed. It involves reordering the line impedance matrix by enforcing a specific permutation pattern while relaxing the others, thereby enabling the linearization of constraints that originally contain quadratic terms.
	
\end{enumerate}

The remainder of this paper is structured as follows. Section II presents  underground cable energization,  distribution transformer energization, and swtichgear phase swapping models. In Section III, the problem is formulated to address the unique characteristics of underground PDS restoration using IBRs. In Section IV, the linearization technique is presented for problem solution. Finally, case studies are discussed in Section V, and in Section VI, the paper is summarized with future directions.

\vspace{-5pt}
\section{Models Related to Underground PDS Restoration}\label{SECII}

In this section, the underground PDS is described. Then, to address the unique challenges in underground PDS restoration using IBRs, three mathematical models are proposed. They are underground cable energization model considering inrush current, distribution transformer energization model considering ferroresonance, phase swapping model for smart switchgears. In the rest of this section, the models are discussed in detail.

\vspace{-5pt}

\subsection{Underground PDS Description}
\begin{figure}[t]
	\centering
	\includegraphics[width=3.2in]{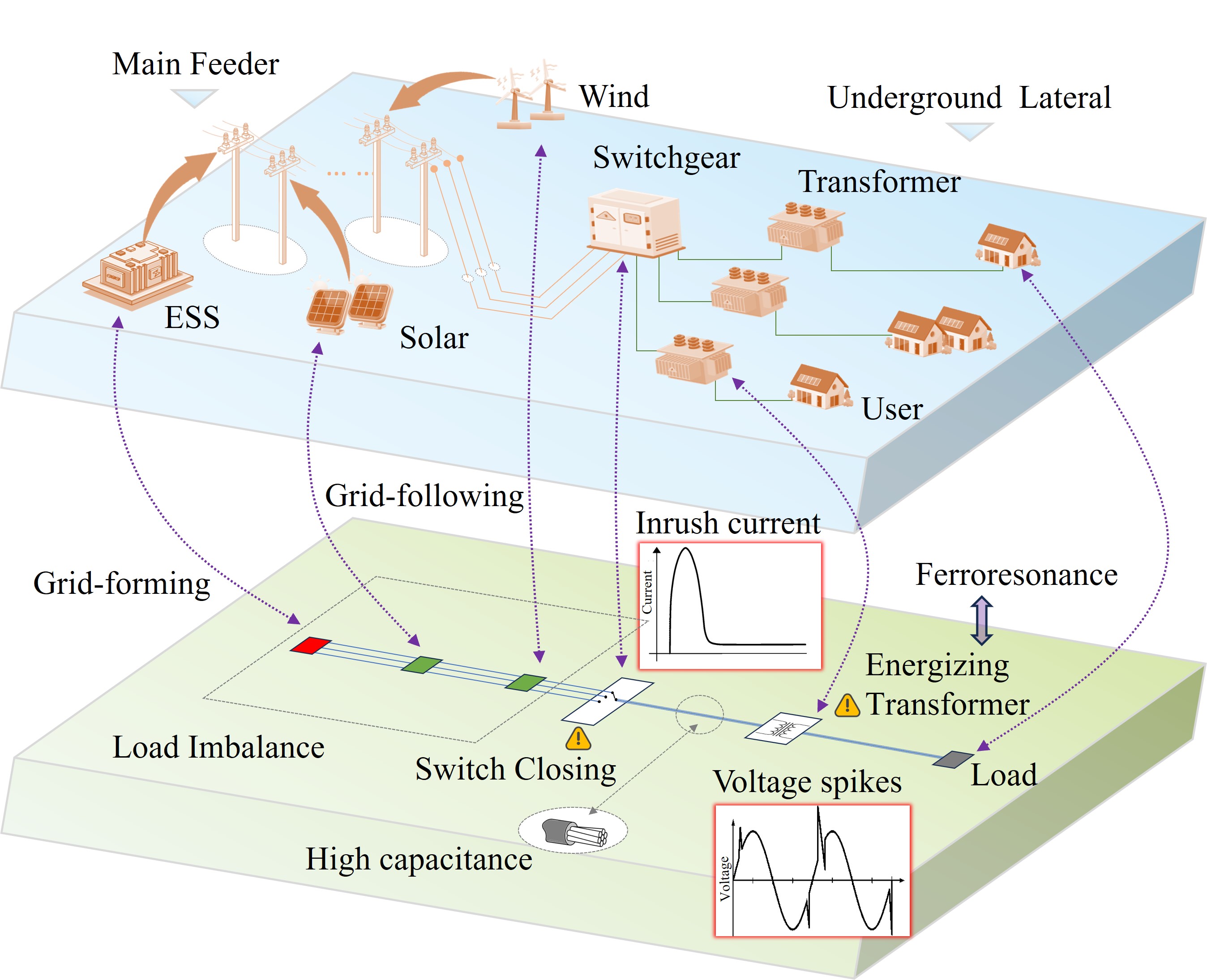}
	\caption{Diagram of underground PDS model description.}
	\label{UPDSModel}
\end{figure}

Fig. \ref{UPDSModel} shows a common configuration of underground PDSs. First of all, the three-phase main feeder, can be underground lines but typically constructed as overhead lines, is responsible for delivering power to the locations near the loads. Herein, we use $i,j\in\mathds{N}$ to represent the index of  nodes, and $(i,j)\in\mathds{L}$ to the index of  lines. For each line, $\boldsymbol{z}_{ij}=\boldsymbol{r}_{ij}+j\boldsymbol{x}_{ij}$ represents the line impedance matrix. Secondly, pad-mounted smart switchgears are responsible to branch off from the main feeder, connecting to the underground laterals, which can be single, two, or three phases. We use $\varphi,\phi\in\{a,b,c\}$ to represent the index of phases. Lastly, pad-mounted distribution transformers step down the voltage level,  completing the service to the end users. For load restoration during blackouts, IBRs with grid-forming and grid following modes are utilized. Specifically, ESSs are operated in grid-forming mode for regulating voltage and frequency. PVs and WTs are operated in grid-following modes, injecting power for restoration by synchronizing with the network references using phase-locked loops. For overhead systems, the restoration in similar configuration has been well investigated. However, when underground cables are involved, the inrush current, ferrorresonance, and load imbalance, shown in Fig. \ref{UPDSModel}, must be controlled and managed.

\vspace{-5pt}

\subsection{Underground Cable Energization Model Considering Inrush Current}

\begin{figure}[t]
	\centering
	\includegraphics[width=3.4in]{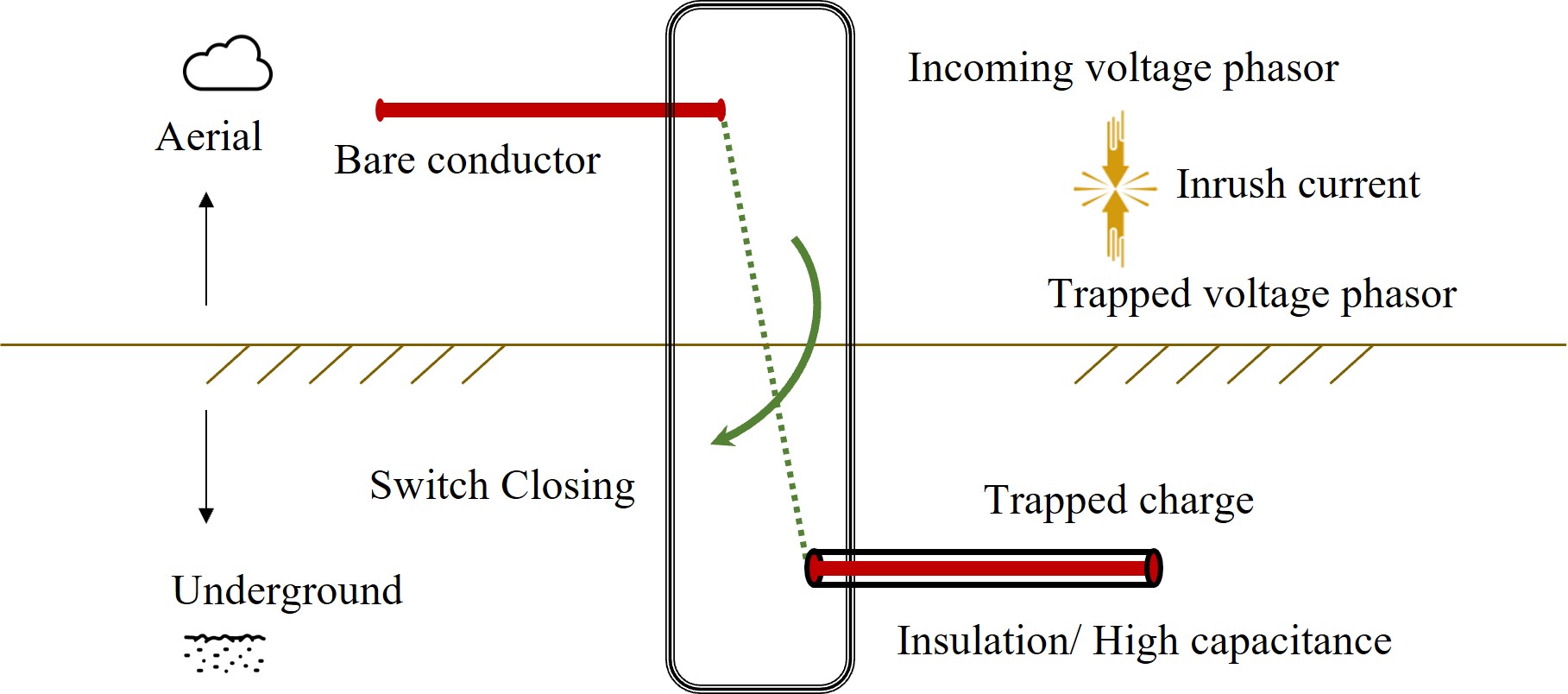}
	\caption{Energizing underground cables considering inrush current.}
	\label{Inrushcurrent}\vspace{-10pt}
\end{figure}

Underground cables are effective in reducing the exposure to extreme weather events. However, the increased capacitance resulting from the smaller spacing between the conductors can lead to trapped charge \cite{pordanjani2016discharge}. This trapped charge can persist for a long time because of the slow discharge rate imposed by the insulation. An illustration is shown in Fig. \ref{Inrushcurrent}. When energizing underground power lines with trapped charge, inrush current can occur, often exceeding the normal current by several times. This inrush current must be managed seriously, as its cumulative impact will degrade or even damage electrical devices. To this end, an inrush current model is proposed to characterize the transient behavior of energizing underground cables. The model aims to quantify the inrush current by considering the voltage difference across the switch in switchgears.

\textit{Theorem}: To capture the phase dynamics, we first use $\Delta \theta_{ij}$ to denote the phase angle difference between the two ends of switch $(i,j)$, where node $i$ is the end connected to the main feeder, and node $j$ is the end  to the underground laterals. Since the switch's impedance is much smaller than the underground cable’s capacitive reactance, it can be neglected. Accordingly, the squared voltage difference across the two ends is
\begin{equation}\label{ICM1}
	\Delta {V}_{ij} = 1/2 (\tilde{V}_{i} - \tilde{V}_{j}^{\text{trap}}) + \Delta \theta_{ij},
\end{equation}
where $\tilde{V}_{j}^{\text{trap}}$ is the squared magnitude of voltage considering trapped charge. Note that, unless otherwise specified, $j$ is used throughout the paper to represent the underground lateral side of switchgears. Also, $\Delta \theta_{ij}$ is a correction term, denoting the phase angle-dependent component between nodes $i$ and $j$. 

\textit{Proof}: Let $V_i = |V_i| e^{j\theta_i}$ and $V_j = |V_j| e^{j\theta_j}$, which represent the voltage phasors at nodes $i$ and $j$, respectively. Then, the voltage difference across the two ends of the switch is
\begin{equation}\label{P1}
	\Delta {V}_{ij}=|V_i| e^{j\theta_i}-|V_j| e^{j\theta_j}.
\end{equation}

By using Euler’s formula, equation (\ref{P1}) can be expanded as 
$
\Delta {V}_{ij}=(|V_i| \cos\theta_i - |V_j| \cos\theta_j) + j (|V_i| \sin\theta_i - |V_j| \sin\theta_j)
$. To mitigate inrush current, the switch is typically closed when the phase-angle difference is very small, commly within $\pm 10^{\circ}$. Accordingly, the following approximation holds: 
$\cos\Delta \theta_{ij} \approx 1-(\Delta \theta_{ij})^2/2$, and $\sin\Delta \theta_{ij} \approx \Delta \theta_{ij}
$. Taking $\theta_j=0$ as the reference, we have $\cos\theta_j=1$. From $\cos \theta_i = \cos\left(\theta_j + \Delta \theta_{ij}\right) \approx \cos \theta_j \cos \Delta \theta_{ij} - \sin \theta_j \sin \Delta \theta_{ij}
$, we have $\cos \theta_i\approx 1$. Then, the real part of $\Delta \theta_{ij}$ becomes
\begin{equation}\label{P2}
	|V_i| \cos \theta_i - |V_j| \cos \theta_j \approx |V_i| - |V_j|.
\end{equation}
where the right hand side is the voltage magnitude difference. Similarly, from $\sin \theta_i = \sin\bigl(\theta_j + \Delta\theta_{ij}\bigr)
\approx \sin \theta_j \cos \Delta\theta_{ij} + \cos \theta_j \sin \Delta\theta_{ij}$, we have $\sin \theta_i\approx \Delta\theta_{ij}$. Then, the imaginary part of $\Delta \theta_{ij}$ becomes
\begin{equation}
	|V_i| \sin \theta_i - |V_j| \sin \theta_j \approx |V_i| \Delta\theta_{ij}.
\end{equation}

Then, the magnitude of voltage difference can be stated as
\begin{equation}\label{P3}
	|\Delta V_{ij}| = \sqrt{\left(|V_i| - |V_j|\right)^2 + \left(|V_i|\,\Delta\theta_{ij}\right)^2}.
\end{equation}

Since switchgear operation and restoration responses occur rapidly, yet the discharging of trapped charge in underground cables is relatively slow, the voltage at the terminal connected to the underground cable will not drop significantly. It means that $|V_i| - |V_j|$ is small. Hence, equation (\ref{P3}) simplifies to
\begin{equation}
	|\Delta V_{ij}| \approx |V_i| - |V_j| + |V_i| \Delta \theta_{ij}.
\end{equation}

Let $\tilde{V}_i=|V_i|^2$ and $\tilde{V}_j=|V_j|^2$, we have $\tilde{V}_i - \tilde{V}_j=(|V_i| - |V_j|)(|V_i| + |V_j|)$. By approximating $|V_i| + |V_j|$ with $2\sqrt{\tilde{V}_{\text{nom}}}$, the following can be derived
\begin{equation}
	|V_i| - |V_j| \approx ({\tilde{V}_i - \tilde{V}_j})/{2\sqrt{\tilde{V}_{\text{nom}}}},
\end{equation}
\begin{equation}
	|V_i| \Delta \theta_{ij} \approx \sqrt{\tilde{V}_{\text{nom}}} \Delta \theta_{ij},
\end{equation}
where $\tilde{V}_{\text{nom}}$ is a constant reference value, e.g., $1$ p.u. \enspace\enspace\enspace\enspace\enspace\enspace\enspace $\square$

Furthermore, based on the squared voltage difference $\Delta {V}_{ij}$, the inrush current peak can be obtained by
\begin{equation}\label{ICM2}
	\mathcal{I}_{ij} = C^\text{tot}\frac{\Delta {V}_{ij}}{\Delta t}, 
\end{equation}
where $C^\text{tot}$ is the  capacitance seen by the swtichgear considering all the downstream underground laterals.

\begin{figure}[t]
	\centering
	\includegraphics[width=3.1in]{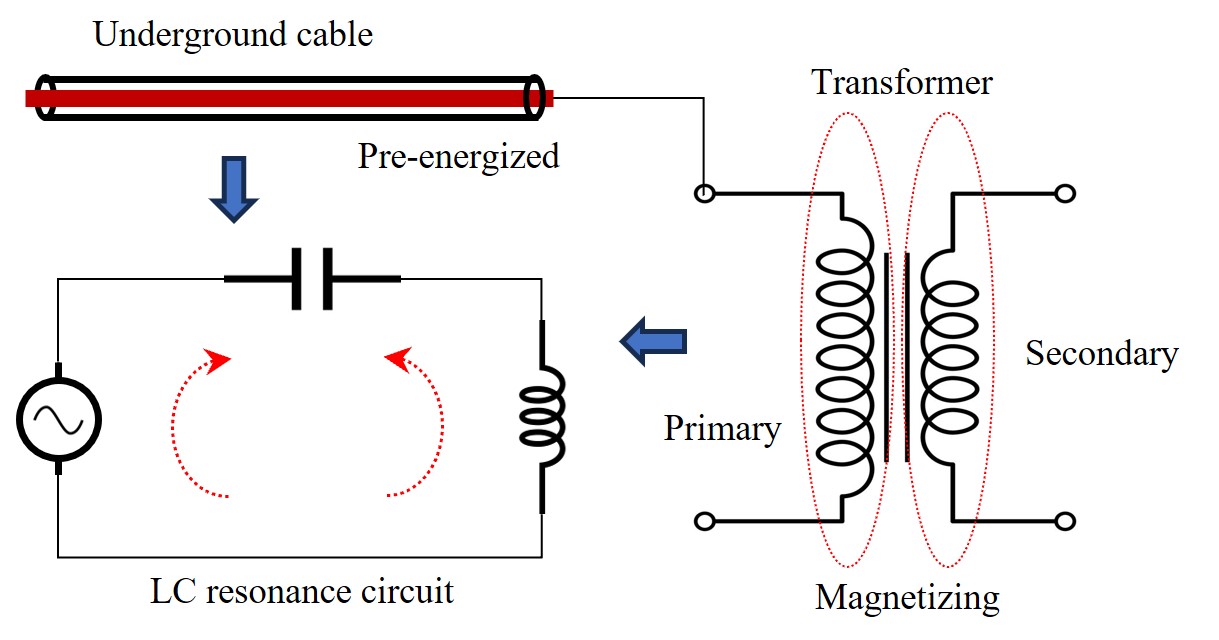}
	\vspace{-5pt}
	\caption{An illustration of distribution transformer energization considering ferroresonance.}
	\label{Ferroresonance}
\end{figure}

\vspace{-10pt}

\subsection{Distribution Transformer Energization Model Considering Ferroresonance} \label{FRM}

Ferroresonance is an oscillatory phenomenon happens when the capacitance of power lines is ignorable. It is much severer when restoring underground PDSs, and energizing distribution transformers by underground cables \cite{mellik2018proactive}. The high capacitance of cables is placed in series with the distribution transformer's primary inductance, creating a resonant circuit. As the transformer is energized, its magnetizing inductance decreases due to the core saturation effect. During this process, the resonant frequency $f^\text{osc}$ can be equaled to the system frequency $f^\text{nom}$ at some points, hence meeting the resonance condition. Subsequently, distorted high voltages will be induced, which may damage distribution transformers.

As shown in Fig. \ref{Ferroresonance}, if we look into the underground lateral from the point of switchgear, the network can be regarded as a series–parallel RLC circuit. Then, the equivalent inductance $L^\text{osc}$ that results in the resonance condition can be obtained by
\begin{equation}
	f^\text{nom}=\frac{1}{2\pi\sqrt{L^\text{osc}C^\text{eq}}}, 
\end{equation}
where $C^\text{eq}$ is the equivalent capacitance. Moreover, to describe the oscillation, the quality factor (Q-factor) is adopted:
\begin{equation}
	\mathcal{Q}=R^\text{dp}\sqrt{\frac{C^\text{eq}}{L^\text{osc}}},
\end{equation}
where $R^\text{dp}$ represents the aggregated damping resistance associated with all loads downstream the switchgear. Physically, a relatively strong damping with lower $R^\text{dp}$ results in a lower Q-factor, meaning that the resonance is less severe. Conversely, a higher $R^\text{dp}$ leads to a higher $\mathcal{Q}$, which increases the risk of sustained overvoltage oscillations. To mitigate the ferroresonance, a practical approach is to maintain the damping level, ensuring sufficient downstream loading during transformer energization. The damping resistance can be estimated  through the ZIP load model \cite{wang2014time,ma2022robust}, which is stated as $R^\text{dp}=|V_j|^2/(Zp^\text{tot})$, considering the impedance part of the load. Herein, $Z$ represent the $Z$ type component, typically ranging from $0.2$ to $0.4$, $p^\text{tot}$ denotes the total active load downstream the switchgear, and $|V_j|^2=\tilde{V}_j$. Except for load demand which contributes to the natural damping, utilities will also use dedicated devices such as pre-insertion resistors and portable protective air gaps for overvoltage control \cite{abdelazim2016case}. 

\begin{figure}[t]
	\centering
	\includegraphics[width=3.4in]{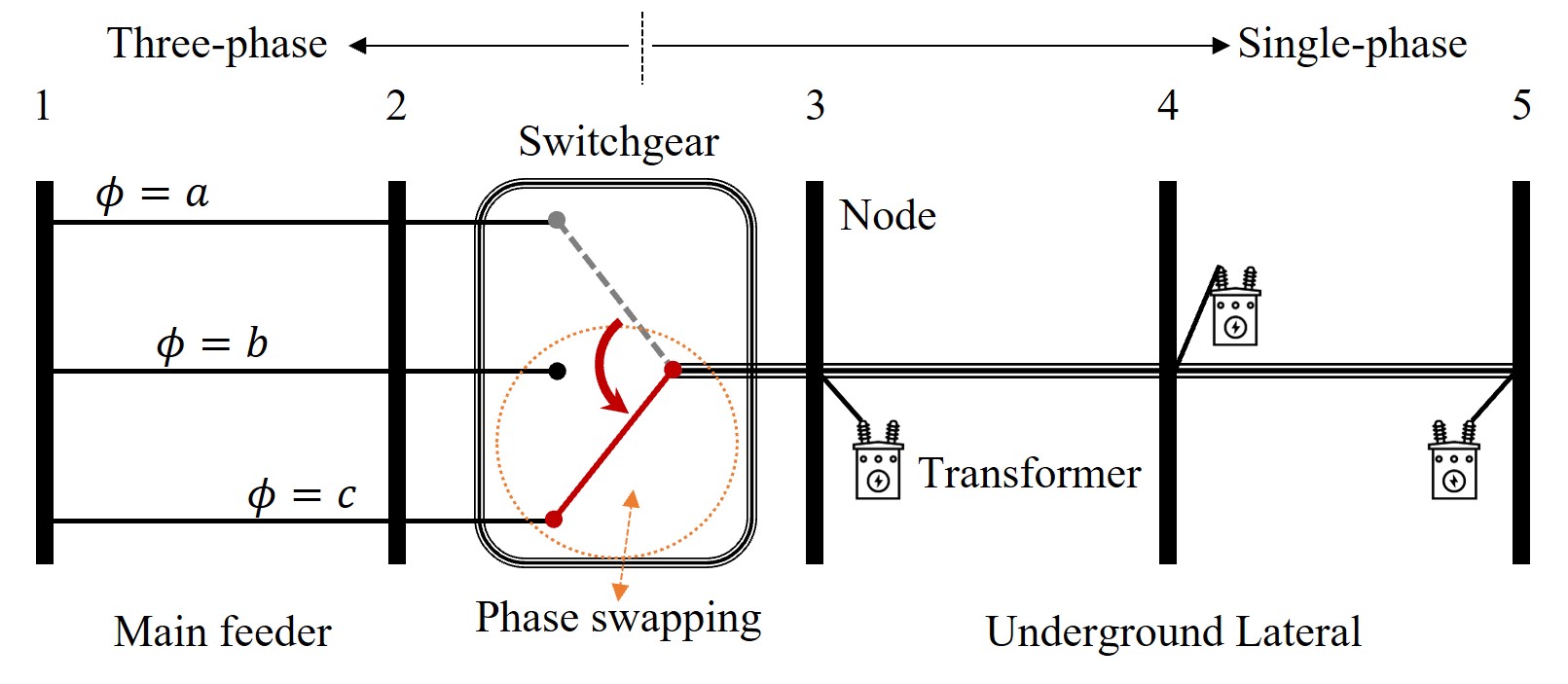}
	\vspace{-5pt}
	\caption{An illustration of phase swapping using swtichgears.}
	\label{Phaseswapping}
\end{figure}

\vspace{-10pt}

\subsection{Phase Swapping Model for Smart Switchgears}\label{PSM}

Phase swapping refers to rearranging the phase connection of laterals with the main feeder dynamically, which is achieved by smart switchgears. An example of phase swapping is shown in Fig. \ref{Phaseswapping}. Initially, a single-phase lateral is connected to phase $a$, while after swapping, this lateral is reconnected to phase $c$. The phase swapping is necessary, since it can address the load imbalance issue, alleviating the stress on IBRs and increasing the overall restored load \cite{zhu1998phase}. For modeling, we first present a $3\times 3$ phase swapping matrix $\boldsymbol{\mathcal{S}}_g$, stated as
\begin{eqnarray}
	\begin{dcases}
		\boldsymbol{\mathcal{S}}_{g}(\varphi,\phi)=1, \enspace \text{if} \enspace \phi\rightarrow \varphi,\\
		\boldsymbol{\mathcal{S}}_{g}(\varphi,\phi)=0, \enspace \text{if} \enspace \phi\nrightarrow \varphi.
	\end{dcases}
\end{eqnarray}

Specifically, when $\boldsymbol{\mathcal{S}}_{g}(\varphi,\phi)=1$, it implies that phase $\phi$ on the underground lateral is reconnected to phase $\varphi$ on the main feeder by operating the switchgear. Note that in the following, we use $\varphi$ to represent the phase on the main feeder side of switchgear $g$, and $\phi$ the phase on the underground lateral side of switchgear $g$. In addition, we use $\boldsymbol{p}_{j,\phi}^{t}$ and $\boldsymbol{q}_{j,\phi}^{t}$ to represent the single-phase active and reactive load demands of node $m$. For example, when $\phi=b$, we have $\boldsymbol{p}_{j,\phi}^{t}=[0,p_{j,b}^{t},0]^\top$ and $\boldsymbol{q}_{j,\phi}^{t}=[0,q_{j,b}^{t},0]^\top$. Accordingly, the load demand after phase swapping can be modeled as
\begin{eqnarray}\label{II1}
	\textstyle\boldsymbol{p}_{j}^{t}=\sum_{\phi\in\Phi}\boldsymbol{\mathcal{S}}_{g}\boldsymbol{p}_{j,\phi}^{t}, \forall j\in\mathds{N}_g,
\end{eqnarray}
\begin{eqnarray}\label{II2}
	\textstyle\boldsymbol{q}_{j}^{t}=\sum_{\phi\in\Phi}\boldsymbol{\mathcal{S}}_{g}\boldsymbol{q}_{j,\phi}^{t}, \forall j\in\mathds{N}_g.
\end{eqnarray}

Equations (\ref{II1})-(\ref{II2}) represent load transferring when conducting phase swapping. They mean that all the loads on the lateral downstream of switchgear $g$ will be rearranged according to phase swapping matrix $\boldsymbol{\mathcal{S}}_g$. An example illustrating the single-phase swapping of phase $b$ is presented in Table \ref{tab:swapping}. For example, when $\boldsymbol{\mathcal{S}}_{g}(1,2)=1$, the load on phase $b$ can be switched from $\boldsymbol{p}_{m,\phi}^{t}=[0,p_{m,b}^{t},0]^\top$ to  $\boldsymbol{p}_{m,\phi}^{t}=[p_{m,b}^{t},0,0]^\top$. It implies that the load on phase $b$ will be transferred to phase $a$. Note that equations (\ref{II1})-(\ref{II2}) can be applied to model two-phase and three-phase swapping.

\begin{table}[t]
	\renewcommand{\arraystretch}{1.3}
	\caption{Single-Phase Swapping of Phase $b$}
	\centering
	\setlength{\tabcolsep}{8pt}
	\begin{tabular}{c c c c}
		\toprule
		\(\boldsymbol{\mathcal{S}}_{g}\) 
		& {Before} 
		& {After} 
		& {Description} \\
		\midrule
		\(\begin{pmatrix}
			0 & 1 & 0\\
			1 & 0 & 0\\
			0 & 0 & 1
		\end{pmatrix}\)
		& \(\begin{pmatrix}
			0\\
			p_{m,b}^t\\
			0
		\end{pmatrix}\)
		& \(\begin{pmatrix}
			p_{m,b}^t\\
			0\\
			0
		\end{pmatrix}\)
		& Transferred to phase \(a\) \\
		\midrule
		\(\begin{pmatrix}
			1 & 0 & 0\\
			0 & 1 & 0\\
			0 & 0 & 1
		\end{pmatrix}\)
		& \(\begin{pmatrix}
			0\\
			p_{m,b}^t\\
			0
		\end{pmatrix}\)
		& \(\begin{pmatrix}
			0\\
			p_{m,b}^t\\
			0
		\end{pmatrix}\)
		& Remain with phase \(b\) \\
		\midrule
		\(\begin{pmatrix}
			1 & 0 & 0\\
			0 & 0 & 1\\
			0 & 1 & 0
		\end{pmatrix}\)
		& \(\begin{pmatrix}
			0\\
			p_{m,b}^t\\
			0
		\end{pmatrix}\)
		& \(\begin{pmatrix}
			0\\
			0\\
			p_{m,b}^t
		\end{pmatrix}\)
		& Transferred to phase \(c\) \\
		\bottomrule
	\end{tabular}
	\label{tab:swapping}
\end{table}

\vspace{-5pt}

\section{Underground PDS Restoration\\Problem Formulation}

In this section, the problem of underground PDS restoration using IBRs  is studied. To address the unique characteristics of underground PDSs, the  models proposed in Section \ref{SECII}, which are related to inrush current, ferroresonance and phase swapping, are included as constraints. The problem is formulated as an MINLP problem with the objective to maximize the load restoration over a multi-period time horizon. In the rest of this section, the problem formulation is presented in detail. 

\vspace{-5pt}

\subsection{Constraints}
To achieve an effective underground PDS restoration using IBRs, the constraints related to the operations of pad-mounted smart switchgears, pad-mounted distribution transformers and IBRs are developed in this subsection.

\subsubsection{Underground Cable Inrush Current Constraints}

In underground PDSs, smart switchgears are capable of monitoring the voltage magnitudes and phase angles across the two terminals of switches. This ensures the switch operation and phase swapping to be implemented as commanded with a minimum inrush current. To achieve this goal, equations (\ref{ICM1}) and (\ref{ICM2}) are included as constraints after modification as follows,
\begin{eqnarray}\label{ICC1}
	\Delta \boldsymbol{\mathcal{S}}^t_{g} = \boldsymbol{\mathcal{S}}^t_{g}-\boldsymbol{\mathcal{S}}^{(t-1)}_{g}, \forall t\geq 1,
\end{eqnarray}
\begin{eqnarray}\label{ICC2}
	-\Delta \boldsymbol{\mathcal{S}}^t_{g}(\varphi,\phi) \leq x_{g,\varphi}^t \leq  \Delta \boldsymbol{\mathcal{S}}^t_{g}(\varphi,\phi)+\epsilon, \forall \phi\in\Phi,
\end{eqnarray}
\begin{eqnarray}\label{ICC3}
	\textstyle1/3\sum\nolimits_{\phi\in\Phi}{x'}_{g,\varphi\phi}^t\leq{x}_{g,\varphi}^t\leq \sum\nolimits_{\phi\in\Phi}{x'}_{g,\varphi\phi}^t, \forall \phi\in\Phi,
\end{eqnarray}
\begin{eqnarray}\label{ICC4}
	\textstyle\mathcal{F}-M(1-x_{g,\varphi}^t)\leq\Delta {V}_{ijk,\varphi}^t \leq \mathcal{F}+M(1-x_{g,\varphi}^t),
\end{eqnarray}
\begin{eqnarray}\label{ICC5}
	-Mx_{g,\varphi}^t\leq \Delta {V}_{ijk,\varphi}^t \leq Mx_{g,\varphi}^t,
\end{eqnarray}
where $x_{g,\varphi}^t$ and ${x'}_{g,\varphi\phi}^t$ are auxiliary variables for representing the time-dependent connection of switchgears. An example of switch operation and phase swapping of phase $\varphi=a$ are listed in Table \ref{Connection}. Specifically, scenario 1 means phase $a$ will remain its connection at time $t$. Scenarios 2 and 3 indicate that phase $a$ is reconnected to phases 2 or 3 at time $t$. Scenario 4 is used to model the switch opening operation of phase $a$. Scenario 5 means phase $a$ will remain its disconnection as the same as time $t-1$. Scenarios 6, 7 and 8 mean that phase $a$ which is disconnected at time $t-1$ will be reconnected to phase $a$, $b$ and $c$, respectively. We can see that all connection operations result in an enabled $x_{g,\varphi}^t=1$. Subsequently, constraints (\ref{ICC4})-(\ref{ICC5}) are adopted to enable inrush current when energizing an underground cable. While when $x_{g,\varphi}^t=0$, it is disabled. Note that $\mathcal{F}=1/2 (\tilde{V}_{ik,\varphi}^t - \sum_{\phi}\tilde{V}_{j,\varphi}^{t,\text{trap}}\boldsymbol{\mathcal{S}}^t_{g}(\varphi,\phi)) + \Delta \theta_{ij,\varphi}^t$, denoting the voltage difference between one single-phase lateral $\phi$ and phase $\varphi$ on the main feeder. Also, $\Delta \theta_{ij,\varphi}^t$ is the phase $\varphi$ angle difference of the switchgear between nodes $i$ and $j$ at time $t$. Note that phase $\varphi$ is on the main feeder side of the switch. 

Furthermore, the inrush current can be obtained by equation (\ref{ICM2}), i.e., $\mathcal{I}_{ijk,\varphi}^t=C\Delta {V}_{ijk,\varphi}^t/\Delta t$. It represents the inrush current of phase $\varphi$ of the main feeder when operating switchgear $g$ between nodes $i$ and $j$. Then, the inrush current should be limited by the following constraint:
\begin{equation}
	\mathcal{I}_{ijk,\varphi}^t \leq \mathcal{I}_{ij}^\text{max}, \forall \varphi\in\Phi, t\in\mathds{T},
\end{equation}
which means the inrush current cannot exceed the maximum current limitation. Note that the value of $\mathcal{I}_{ij}^\text{max}$ can be determined by the maximum allowable inrush current.

\begin{table}[t]
	\centering
	\setlength{\tabcolsep}{8.8pt}
	\caption{An Example of Connection of Switchgears}
	\begin{tabular}{c |c |c |c |c }
		\toprule
		Scenario & 1 & 2 & 3 & 4  \\
		\midrule
		\(\boldsymbol{\mathcal{S}}_{g}^t(\varphi,\cdot)\)  
		& \([1,0,0]\) 
		& \([0,1,0]\) 
		& \([0,0,1]\) 
		& \([0,0,0]\)  \\[3pt]
		\(\Delta\boldsymbol{\mathcal{S}}_g^t(\varphi,\cdot)\)  
		& \([0,0,0]\) 
		& \([-1,1,0]\) 
		& \([-1,0,1]\) 
		& \([-1,0,0]\)  \\[3pt]
		\({x'}_{g,\varphi\phi}^t\)  
		& \([0,0,0]\) 
		& \([0,1,0]\) 
		& \([0,0,1]\) 
		& \([0,0,0]\)  \\[3pt]
		\(x_{g,\varphi}^t\) 
		& 0 & 1 & 1 & 0  \\
		\midrule\midrule
		Scenario & 5 & 6 & 7 & 8  \\
		\midrule
		\(\boldsymbol{\mathcal{S}}_{g}^t(\varphi,\cdot)\)  
		& \([0,0,0]\) 
		& \([1,0,0]\) 
		& \([0,1,0]\) 
		& \([0,0,1]\)  \\[3pt]
		\(\Delta\boldsymbol{\mathcal{S}}_g^t(\varphi,\cdot)\)  
		& \([0,0,0]\) 
		& \([1,0,0]\) 
		& \([0,1,0]\) 
		& \([0,0,1]\)  \\[3pt]
		\({x'}_{g,\varphi\phi}^t\)  
		& \([0,0,0]\) 
		& \([1,0,0]\) 
		& \([0,1,0]\) 
		& \([0,0,1]\)  \\[3pt]
		\(x_{g,\varphi}^t\) 
		& 0 & 1 & 1 & 1  \\
		\bottomrule
	\end{tabular}
	\label{Connection}
	\begin{flushleft} Note: For scenarios 1 to 4, we have  \(\boldsymbol{\mathcal{S}}_{g}^{(t-1)}(\varphi,\cdot)=[1,0,0]\),  for scenarios 5 to 8, we have  \(\boldsymbol{\mathcal{S}}_{g}^{(t-1)}(\varphi,\cdot)=[0,0,0]\).
	\end{flushleft}\vspace{-10pt}
\end{table}

\subsubsection{Distribution Transformer Energization Constraints}

To effectively manage the ferroresonance phenomenon when energizing distribution transformers using energized underground cables, the model proposed in Subsection \ref{FRM} is incorporated as constraints in the optimization framework.
\begin{eqnarray}\label{DTEC1}
	\mathcal{Q}_g^t=R_g^{t,\text{dp}}\sqrt{\frac{C_g^\text{eq}}{L_g^\text{osc}}}, \forall g\in\mathds{G}, t\in\mathds{T},
\end{eqnarray}
\begin{eqnarray}\label{DTEC2}
	\mathcal{Q}_g^t - \mathcal{Q}_g^\text{max}\leq  M(1-\alpha_{g}^t), \forall g\in\mathds{G}, t\in\mathds{T},
\end{eqnarray}
\begin{eqnarray}\label{DTEC3}
	\alpha_{g}^{t}+\beta_g^{t-1}\geq \beta_g^t, \forall g\in\mathds{G}, t\in\mathds{T}.
\end{eqnarray}

Constraint (\ref{DTEC1}) is used to calculate the Q-factor downstream switchgear $g$ at time $t$ by considering time-dependent damping resistance $R_g^{t,\text{dp}}$, which denotes a continuous variable since it is related to $\tilde{V}_j$. Constraints (\ref{DTEC2})-(\ref{DTEC3}) mean a switchgear can be closed to energize the downstream transformers, only when the Q-factor is lower than a threshold. Notice that constraints (\ref{DTEC1})-(\ref{DTEC2}) are applied to all three phases of switchgears,  which are allowed to operate only if the Q-factor condition is satisfied for all phases. Constraint (\ref{DTEC3}) means that the ferroresonance condition will be checked every time when closing switchgears.

\subsubsection{Smart Switchgear Phase Swapping Constraints}

In subsection \ref{PSM}, equations (\ref{II1})-(\ref{II2}) provides the model for phase swapping using smart switchgears. Except for modeling, the swapping  must be conducted under the following constraints:
\begin{eqnarray}\label{PSC1}
	\textstyle 0\leq\sum_{\varphi\in\Phi}\boldsymbol{\mathcal{S}}^t_g(\varphi,\phi)\leq 1, \forall \phi\in\Phi,
\end{eqnarray}
\begin{eqnarray}\label{PSC2}
	\textstyle 0\leq\sum_{\phi\in\Phi}\boldsymbol{\mathcal{S}}^t_g(\varphi,\phi)\leq 1, \forall \varphi\in\Phi,
\end{eqnarray}
\begin{eqnarray}\label{PSC3}
	\textstyle 3\beta_g^t\leq\sum_{\varphi,\phi\in\Phi}\boldsymbol{\mathcal{S}}^t_g(\varphi,\phi)\leq 3\beta_g^t,\forall g\in\mathds{G}.
\end{eqnarray}

Constraint (\ref{PSC1}) ensures that one single-phase lateral can be reconnected to a specific phase on the main feeder. 
Constraint (\ref{PSC2}) implies that one phase $\varphi$ on the main feeder can accommodate only one phase of lateral after reconnection. Constraint (\ref{PSC3}) implies the opening operation must be performed consistently across all three phases of a switchgear.

\subsubsection{Grid-Forming ESS Constraints} To achieve an effective operation of grid-forming ESSs, the constraints related to ESS charging and discharging dynamics are developed:
\begin{eqnarray}\label{ESSC1}
	{c}_{k}^{t}+{d}_{k}^{t}\leq 1, \forall k\in\mathds{K}, t\in\mathds{T},
\end{eqnarray}
\begin{eqnarray}\label{ESSC2}
	\textstyle \mathcal{E}_{k}^{t}=\mathcal{E}_{k}^{t-1}+\{\eta_k^{ch}(\boldsymbol{1}^\top\boldsymbol{{C}}_{k}^{t})-1/\eta_k^{dis}(\boldsymbol{1}^\top\boldsymbol{{D}}_{k}^{t})\}\Delta t,
\end{eqnarray}
\begin{eqnarray}\label{ESSC3}
	\boldsymbol{P}_{k}^{t}=\boldsymbol{{C}}_{k}^{t}-\boldsymbol{{D}}_{k}^{t}, \forall k\in\mathds{K},
\end{eqnarray}
\begin{eqnarray}\label{ESSC4}
	{c}_{k}^{t}{\boldsymbol{{C}}}_{k}^{\text{min}}\leq\boldsymbol{{C}}_{k}^{t}\leq{c}_{k}^{t}{\boldsymbol{{C}}}_{k}^{\text{max}}, \forall k\in\mathds{K},
\end{eqnarray}
\begin{eqnarray}\label{ESSC5}
	{d}_{k}^{t}{\boldsymbol{{D}}}_{k}^{\text{min}}\leq\boldsymbol{{D}}_{k}^{t}\leq{d}_{k}^{t}{\boldsymbol{{D}}}_{k}^{\text{max}}, \forall k\in\mathds{K},
\end{eqnarray}
\begin{eqnarray}\label{ESSC6}
	\boldsymbol{{Q}}_{k}^{\text{min}}\leq\boldsymbol{Q}_{k}^{t}\leq\boldsymbol{{Q}}_{k}^{\text{max}}, \forall k\in\mathds{K},
\end{eqnarray}
\begin{eqnarray}\label{ESSC7}
	\textstyle \mathcal{E}_{k}^{t=0}=\eta_k^\text{ini}\mathcal{E}_{k}^\text{max}, \forall k\in\mathds{K},
\end{eqnarray}
\begin{eqnarray}\label{ESSC8}
	\textstyle\eta_k^\text{min}{{\mathcal{E}}}_{k}^\text{max}\leq \mathcal{E}_{k}^{t}\leq\textstyle\eta_k^\text{max}{{\mathcal{E}}}_{k}^\text{max}, \forall k\in\mathds{K}.
\end{eqnarray}

Constraint (\ref{ESSC1}) means ESSs can operate in either charging state or discharging state. Constraints (\ref{ESSC2})-(\ref{ESSC3}) are used for energy balance considering discharging and charging efficiency. Constraints  (\ref{ESSC4})-(\ref{ESSC5}) ensure that the charging and discharging power at time $t$ are within limitations. Constraint (\ref{ESSC6}) restricts the maximum and minimum reactive power support that ESS $k$ can provide. Constraints (\ref{ESSC7})-(\ref{ESSC8}) set up the initial SOC and limit the minimum and maximum SOC, respectively.

\subsubsection{Grid-Following PV and WT Constraints} PVs and WTs can inject  renewable power into the grid for restoration when bulk systems are unavailable. While, when considering the output fluctuation and limited reactive power support capability \cite{wang2014coordinated}, the following constraints must be satisfied:
\begin{eqnarray}\label{RESC1}
	Pr(\boldsymbol{P}_{j}^{t,\text{RES}}\leq \overline{\boldsymbol{P}}_{j}^{t,\text{RES}}) \geq \tau_j, \forall j\in \mathds{Z}, t\in \mathds{T},
\end{eqnarray}
\begin{eqnarray}\label{RESC2}
	\underline{\boldsymbol{Q}}_{j}^{t,\text{RES}}\leq\boldsymbol{Q}_{j}^{t,\text{RES}}\leq\overline{\boldsymbol{Q}}_{j}^{t,\text{RES}}, \forall j\in \mathds{Z}, t\in \mathds{T}.
\end{eqnarray}

Constraint (\ref{RESC1}) is a chance constraint considering renewable generation uncertainties \cite{shi2024dynamic}. It implies that the probability of the actual power output not exceeding the predicted value is at least $\tau_j$ (confidence level). By assuming that the uncertainty in renewable output is normally distributed, the inverse of the standard normal cumulative distribution can be applied. Then, constraint (\ref{RESC1}) can be linearized as the following,
\begin{eqnarray}\label{RESC3}
	0\leq \boldsymbol{P}_{j}^{t,\text{RES}}  (1+\sigma_j\Phi^{-1}(1-\tau_j)), \forall j\in \mathds{Z}, t\in \mathds{T},
\end{eqnarray}
where $\sigma_j$ is the deviation representing the uncertainty level of the renewable generation at node $j$. And, $\Phi^{-1}(1-\tau_j)$ is the quantile of the standard normal distribution corresponding to  $(1-\tau_j)$. Note that constraint (\ref{RESC3}) reflects a risk-averse formulation, which conservatively reduces the available renewable output considering uncertainties. In addition, constraint (\ref{RESC2}) limits the reactive power support from PVs and WTs according to their designed capacities.

\subsubsection{Topology Constraints} During blackstart restoration, the network is divided into several microgrids. It involves topology reconfiguration by using sectionalizing and tie switches \cite{wang2015self}. One essential group of constraints is for node activation:
\begin{eqnarray}\label{TC1}
	u_{ik} = 1, \forall i=k, k\in\mathds{K},
\end{eqnarray}
\begin{eqnarray}\label{TC2}
	u_{jk}\leq u_{ik}, \forall i,j\in\mathds{N}, k\in\mathds{K},
\end{eqnarray}
\begin{eqnarray}\label{TC3}
	\textstyle \sum_{k\in\mathds{K}}u_{ik}\leq{1}, \forall i\in \mathds{N},
\end{eqnarray}
\begin{eqnarray}\label{TC4}
	u_{ik} + u_{jk} \leq 1+\gamma_{ij}, \forall (i,j)\in\mathds{S},k\in\mathds{K},
\end{eqnarray}
\begin{eqnarray}\label{TC5}
	u_{ik} - u_{jk} \leq 1-\gamma_{ij}, \forall (i,j)\in\mathds{S},k\in\mathds{K},
\end{eqnarray}
\begin{eqnarray}\label{TC6}
	u_{ik}-u_{jk} \geq -(1-\gamma_{ij}), \forall (i,j)\in\mathds{S},k\in\mathds{K}.
\end{eqnarray}

Constraint (\ref{TC1}) enables the  grid-forming ESSs for initiating  restoration. Constraint (\ref{TC2}) means a node can be activated only when its parent node with respect to the same ESS node is activated. Constraint (\ref{TC3}) avoids the situation that one node is covered by more than one grid-forming ESS for voltage and frequency regulation. Constraints (\ref{TC4})-(\ref{TC6}) are used to model the switch operation. If a switch is disconnected, its two ends $i$ and $j$ can not be activated by the same ESS, and vice versa. If line $(i,j)$ is not switchable, the activation status of nodes $i$ and $j$ must always remain consistent. 

To maintain the network radiality when operating switches, the following constraints based on the single-commodity flow and fictitious graph are applied \cite{wang2020radiality}:
\begin{eqnarray}\label{TC7}
	\textstyle\sum\nolimits_{(i,j)\in\mathds{L} }\gamma_{ij} = |\mathds{N}| - |\mathds{K}|,
\end{eqnarray}
\begin{eqnarray}\label{TC8}
	\textstyle\sum\nolimits_{i \in \pi_j} f_{ji} - \sum\nolimits_{i \in \pi_j} f_{ij} = -1,  \forall j \notin \mathds{K},
\end{eqnarray}
\begin{eqnarray}\label{TC9}
	\textstyle\sum\nolimits_{i \in \pi_k} f_{ki}  \geq 1,  \forall k \in \mathds{K},
\end{eqnarray}
\begin{eqnarray}\label{TC10}
	-\Gamma_{ij}M \leq f_{ij}\leq \Gamma_{ij}M, \forall (i,j)\in\mathds{S},
\end{eqnarray}
\begin{eqnarray}\label{TC11}
	\gamma_{ij} \leq \Gamma_{ij}, \forall (i,j)\in\mathds{S}.
\end{eqnarray}

Constraint (\ref{TC7}) defines the number of connected lines in the graph based on the number of grid-forming ESSs. Constraints (\ref{TC8})-(\ref{TC9}) enforce one unit of fictitious commodity flows from  ESS nodes to each other node to  ensure the radial connectivity. Constraints (\ref{TC10})-(\ref{TC11}) are utilized for mapping the distribution network and fictitious network in terms of switch operation.

\subsubsection{Three-Phase Unbalanced Power Flow Constraints} To evaluate the three-phase power flow in underground PDSs, the branch flow model is adopted. The angle
and convex relaxation techniques are applied to
relax the model into a conic problem \cite{farivar2013branch}. Then, the power flow constraints are given by 
\begin{eqnarray}\label{PF1}
	\textstyle\sum_{j'\in\mathds{C}_j}\boldsymbol{P}_{jj'k}^{t}=\boldsymbol{P}_{ijk}^{t}-\boldsymbol{p}_{j}^{t}+\boldsymbol{P}_{j}^{t,\text{RES}}-\boldsymbol{r}_{ij}{\tilde{\boldsymbol{I}}_{ijk}^{t}}+\dot{\boldsymbol{p}}_{ijk}^{t},
\end{eqnarray}
\begin{eqnarray}\label{PF2}
	\textstyle\sum_{j'\in\mathds{C}_j}\boldsymbol{Q}_{jj'k}^{t}=\boldsymbol{Q}_{ijk}^{t}-\boldsymbol{q}_{j}^{t}+\boldsymbol{Q}_{j}^{t,\text{RES}}-\boldsymbol{x}_{ij}{\tilde{\boldsymbol{I}}_{ijk}^{t}}+\dot{\boldsymbol{q}}_{ijk}^{t},
\end{eqnarray}
\begin{eqnarray}\label{PF3}
	{\tilde{\boldsymbol{V}}_{ik}^{t}}-{\tilde{\boldsymbol{V}}_{jk}^{t}}=2(\hat{\boldsymbol{r}}_{ij}\boldsymbol{P}_{ijk}^{t}+\hat{\boldsymbol{x}}_{ij}\boldsymbol{Q}_{ijk}^{t})+\hat{\boldsymbol{z}}_{ij}{\tilde{\boldsymbol{I}}_{ijk}^{t}}+\dot{\boldsymbol{v}}_{jk}^{t},
\end{eqnarray}  
\begin{eqnarray}\label{PF4}
	\tilde{\boldsymbol{I}}_{ijk}^{t}\odot{\tilde{\boldsymbol{V}}_{jk}^{t}}\geq(\boldsymbol{P}_{ijk}^{t})^{\odot 2}+(\boldsymbol{Q}_{ijk}^{t})^{\odot 2}.
\end{eqnarray}

Constraints (\ref{PF1})-(\ref{PF2}) are used for active and reactive power balance. Constraint (\ref{PF3}) is used to calculate the nodal voltage drop, where $\hat{\boldsymbol{r}}_{ij}$, $\hat{\boldsymbol{x}}_{ij}$, and $\hat{\boldsymbol{z}}_{ij}$ represent the equivalent resistance, reactance, and impedance matrices, respectively \cite{shen2020distributed}. Constraint (\ref{PF4}) is a second-order cone constraint, representing the angle and convex relaxation. Note that $\dot{\boldsymbol{p}}_{ijk}^{t}$, $\dot{\boldsymbol{q}}_{ijk}^{t}$,  $\dot{\boldsymbol{v}}_{jk}^{t}$ are slack variables, such that constraints (\ref{PF1})-(\ref{PF3}) can be feasible, when $\boldsymbol{P}_{ijk}^{t}$, $\boldsymbol{Q}_{ijk}^{t}$, and $\tilde{\boldsymbol{V}}_{jk}^{t}$ are forced to $0$. The slack  will be enabled when $u_{jk}=0$ by the following constraints: 
\begin{eqnarray}\label{PF5}
	\boldsymbol{0}\leq{\dot{\boldsymbol{ p}}_{ijk}^t},{\dot{\boldsymbol{ q}}_{ijk}^t}\leq ({1}-u_{jk})\boldsymbol{M}, \forall (i,j)\in\mathds{S},
\end{eqnarray}
\begin{eqnarray}\label{PF5}
	\boldsymbol{0}\leq{\dot{\boldsymbol{v}}_{jk}^t}\leq ({1}-u_{jk})\boldsymbol{M}, \forall j\in\mathds{C}_i.
\end{eqnarray}

\subsubsection{Operational Constraints}
To ensure an effective restoration, the following operational constraints are applied:
\begin{eqnarray}\label{OC1}
	-u_{jk}{\boldsymbol{P}}_k^t \leq \boldsymbol{P}_{ijk}^{t} \leq u_{jk}{\boldsymbol{P}}_k^t, \forall j\in\mathds{C}_i, k\in\mathds{K},
\end{eqnarray}
\begin{eqnarray}\label{OC2}
	-u_{jk}{\boldsymbol{Q}}_k^t \leq \boldsymbol{Q}_{ijk}^{t} \leq u_{jk}{\boldsymbol{Q}}_k^t, \forall j\in\mathds{C}_i, k\in\mathds{K},
\end{eqnarray}
\begin{eqnarray}\label{OC3}
	\textstyle\sum\nolimits_{k\in\mathds{K}}u_{ik}\underline{\boldsymbol{V}}\leq\sum\nolimits_{k\in\mathds{K}}\tilde{\boldsymbol{V}}_{ik}^t\leq \sum\nolimits_{k\in\mathds{K}}u_{ik}\overline{\boldsymbol{V}}, \forall t\in\mathds{T},
\end{eqnarray}
\begin{eqnarray}\label{OC4}
	0\leq\tilde{\boldsymbol{I}}_{ijk}^{t}\leq u_{jk}\overline{\boldsymbol{I}}_{ij}, \forall j\in\mathds{C}_i, k\in\mathds{K}.
\end{eqnarray}

Constraints (\ref{OC1})-(\ref{OC2}) ensure the active and reactive power output of ESS $k$ are limited by the rated capacities. Constraint (\ref{OC3}) means the nodal voltages should be within an acceptable range. Constraint (\ref{OC4}) ensures that the current does not exceed the maximum current capacity. Note that these constraints also force the power flows, voltages and currents to 0, if node $j$ is not supplied by the microgrid established by ESS $k$.

\vspace{-10pt}

\subsection{Objective Function}

The underground PDS restoration problem is formulated as an MINLP problem. The objective is to maximize the multi-period total weighted restored load, given by
\begin{eqnarray}\label{D1}
	\max_{}\sum\limits_{t\in \mathds{T}}\Big\{\sum\limits_{i\in \mathds{N}}\sum\limits_{\phi\in\Phi}w_i \boldsymbol{p}_{i}^{t}-\sum_{(i,j)\in\mathds{L}}\sum_{ \phi\in\Phi}\sum_{k\in\mathds{K}}\boldsymbol{r}_{ij}\tilde{\boldsymbol{I}}_{ijk}^{t}\Big\}.
\end{eqnarray}

The first term in the bracket is the weighted restored load among all phases at time $t$. And, the second term is the total power losses function, which is added to ensure the exactness of the convex relaxation for the second-order cone constraint (\ref{PF4})  \cite{nejad2019distributed}. In addition, the primary variables for underground PDS restoration are explicitly listed as following:
\begin{enumerate}
	
	\item
	$\Delta \theta_{ij,\varphi}^t$: Phase angle differences. These variables, serving as control parameters, will be sent to smart switchgears. The closing operation is conducted automatically when the voltage and phase angles meet the commands. After optimization, the inrush current is minimized. 
	
	\item
	$\alpha_g^t$: Distribution transformer energizing indicators. They show the appropriate time of closing switchgears to energize the primary sides of downstream transformers that satisfy Q-factor conditions, mitigating ferroresonance.
	
	\item $\boldsymbol{\mathcal{S}}^t_{g}$: Time-dependent phase swapping matrix. These variables inform the system operator the phase reconnection decisions for smart switchgear operation at each time $t$. The optimal   $\boldsymbol{\mathcal{S}}^t_{g}$ can mitigate the stress resulted by load imbalance on IBRs with low inertia.
	
\end{enumerate}
	
Except for the primary variables, there are other important variables related to IBR dispatch $\{\boldsymbol{C}_{k}^{t},\boldsymbol{D}_{k}^{t},\boldsymbol{Q}_{k}^{t}\}$ and topology reconfiguration  $\{u_{ik},\gamma_{ij},\beta_g^t\}$, that are determined through the optimization of the underground PDS restoration problem.

\vspace{-5pt}

\begin{figure}[t]
	\centering
	\includegraphics[width=3.4in]{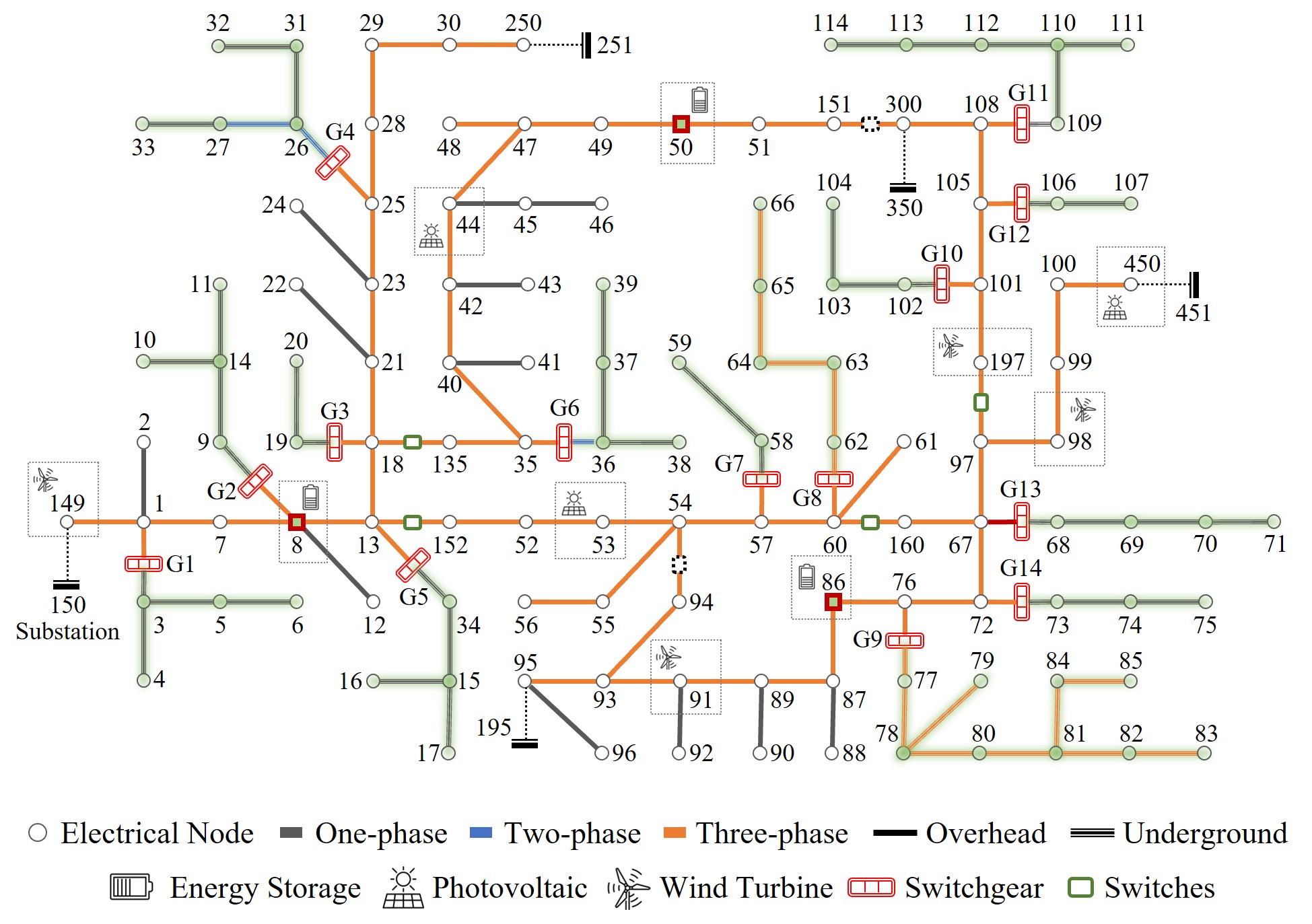}
	\caption{An illustration of the selected underground PDS based on IEEE 123-Node Test Feeder.}
	\label{IEEE123}\vspace{-5pt}
\end{figure}

\begin{figure*}[t]
	\centering
	\includegraphics[width=7in]{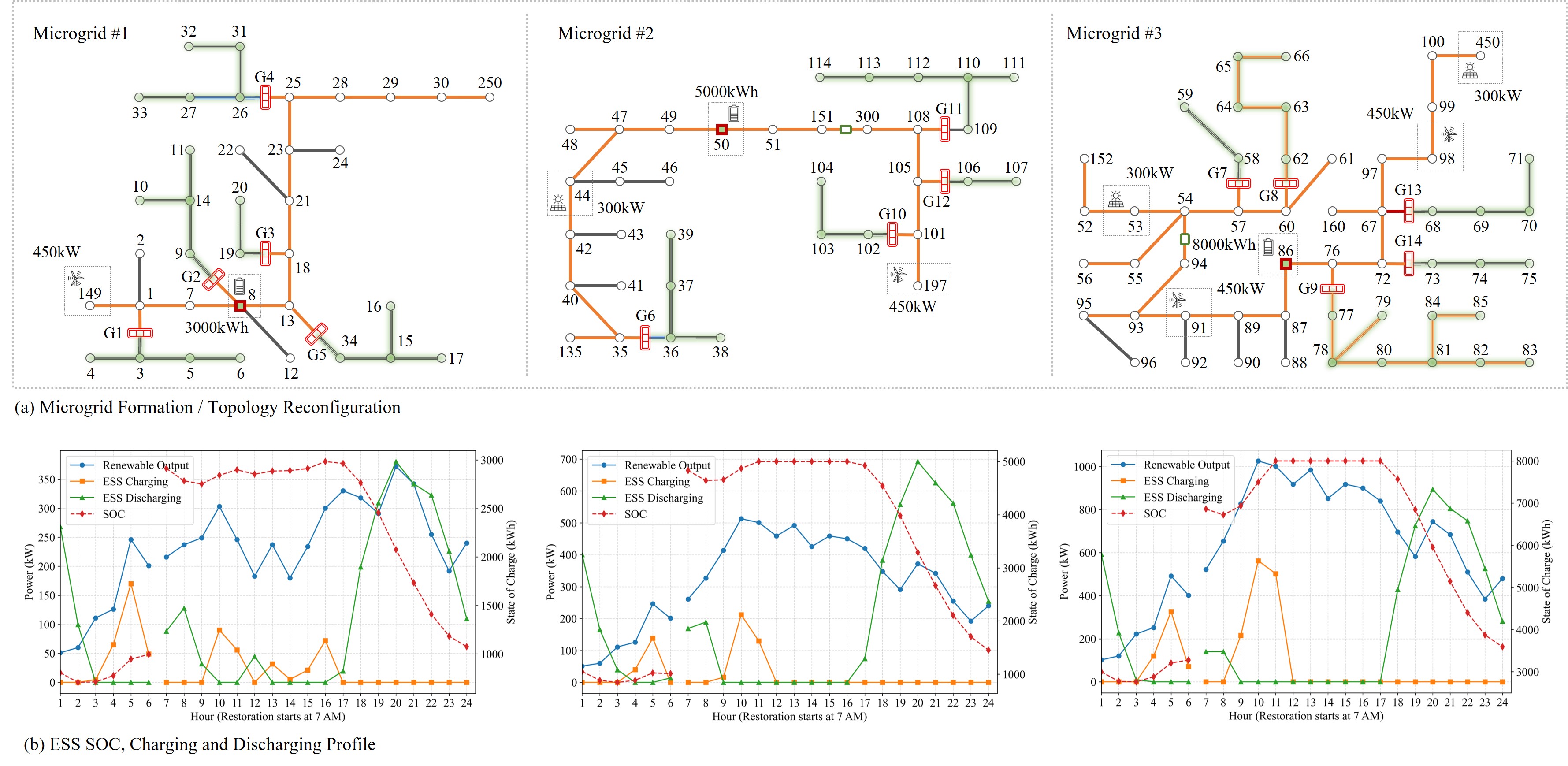}
	\caption{An illustration of the overall restoration performance.}
	\label{Recfig}
\end{figure*}

\section{Underground PDS Restoration\\ Problem Solution}

When including the phase swapping model for switchgears into the optimization, the three-phase unbalanced power flow constraints (\ref{PF1})-(\ref{PF3}) are nonlinear due to the quadratic terms $\boldsymbol{r}_{ij}{\tilde{\boldsymbol{I}}_{ijk}^{t}}$, $\boldsymbol{x}_{ij}{\tilde{\boldsymbol{I}}_{ijk}^{t}}$ and $\hat{\boldsymbol{z}}_{ij}{\tilde{\boldsymbol{I}}_{ijk}^{t}}$. The reason is the order of phases is changed, therefore even though the line configuration is not physically altered, the line impedance matrix $\boldsymbol{z}_{ij}=\boldsymbol{r}_{ij}+j\boldsymbol{x}_{ij}$ should be reordered. This reordering can be expressed as
\begin{eqnarray}
	{\boldsymbol{z}'}_{ij}^t=\boldsymbol{\mathcal{Z}}_g^t\boldsymbol{z}_{ij}^t{\boldsymbol{\mathcal{Z}}_g^t}^\intercal,
\end{eqnarray}
where $\boldsymbol{\mathcal{Z}}_g$ is a permutation matrix indicating the reorder, which has six variations considering three-phase unbalanced systems:
\begin{eqnarray}
	{\boldsymbol{\mathcal{Z}}}_{g,1}^t=\begin{bmatrix}
		1& 0 & 0\\
		0& 1 & 0\\
		0& 0& 1
	\end{bmatrix}
	,
	{\boldsymbol{\mathcal{Z}}}_{g,2}^t=\begin{bmatrix}
		1& 0 & 0\\
		0& 0 & 1\\
		0& 1& 0
	\end{bmatrix},
	\nonumber
\end{eqnarray}
\begin{eqnarray}
	{\boldsymbol{\mathcal{Z}}}_{g,3}^t=\begin{bmatrix}
		0& 1 & 0\\
		1& 0 & 0\\
		0& 0& 1
	\end{bmatrix},	{\boldsymbol{\mathcal{Z}}}_{g,4}^t=\begin{bmatrix}
	0& 1 & 0\\
	0& 0 & 1\\
	1& 0& 0
	\end{bmatrix}
	,\nonumber
\end{eqnarray}
\begin{eqnarray}
	{\boldsymbol{\mathcal{Z}}}_{g,5}^t=\begin{bmatrix}
		0& 0 & 1\\
		1& 0 & 0\\
		0& 1& 0
	\end{bmatrix}
	,
	{\boldsymbol{\mathcal{Z}}}_{g,6}^t=\begin{bmatrix}
		0& 0 & 1\\
		0& 1 & 0\\
		1& 0& 0
	\end{bmatrix}.\nonumber
\end{eqnarray}

If $\boldsymbol{z}_{ij}$ is defined  conventionally with sequence $a$, $b$, $c$, after reordering the self and mutual impedances in the matrix, there are also six variations correspondingly, given by
\begin{equation}
	{\boldsymbol{z}'}_{ij,1}^t=\begin{bmatrix}
		z_{aa} & z_{ab} & z_{ac}\\
		z_{ab} & z_{bb} & z_{bc}\\
		z_{ac}& z_{bc} & z_{cc}
	\end{bmatrix}
	,
	{\boldsymbol{z}'}_{ij,2}^t=\begin{bmatrix}
		z_{aa} & z_{ac} & z_{ab}\\
		z_{ac} & z_{cc} & z_{bc}\\
		z_{ab}& z_{bc} & z_{bb}
	\end{bmatrix}
	,\nonumber
\end{equation}
\begin{equation}
	{\boldsymbol{z}'}_{ij,3}^t=\begin{bmatrix}
		z_{bb} & z_{ab} & z_{bc}\\
		z_{ab} & z_{aa} & z_{ac}\\
		z_{bc}& z_{ac} & z_{cc}
	\end{bmatrix},
	{\boldsymbol{z}'}_{ij,4}^t=\begin{bmatrix}
		z_{bb} & z_{bc} & z_{ab}\\
		z_{bc} & z_{cc} & z_{ac}\\
		z_{ab}& z_{ac} & z_{aa}
	\end{bmatrix}
	,\nonumber
\end{equation}
\begin{equation}
	{\boldsymbol{z}'}_{ij,5}^t=\begin{bmatrix}
		z_{cc} & z_{ac} & z_{bc}\\
		z_{ac} & z_{aa} & z_{ab}\\
		z_{bc}& z_{ab} & z_{bb}
	\end{bmatrix}
	,
	{\boldsymbol{z}'}_{ij,6}^t=\begin{bmatrix}
		z_{cc} & z_{bc} & z_{ac}\\
		z_{bc} & z_{bb} & z_{ab}\\
		z_{ac}& z_{ab} & z_{aa}
	\end{bmatrix}.\nonumber
\end{equation}

Accordingly, a linearization technique is proposed. Specifically, we can observe that if $\boldsymbol{\mathcal{S}}_g^t$ equals to one of the variations of $\boldsymbol{\mathcal{Z}}_{g,v}$, i.e., $\boldsymbol{\mathcal{S}}_g^t=\boldsymbol{\mathcal{Z}}_{g,v}^t$, then $\{\boldsymbol{\mathcal{S}}_g^t-\boldsymbol{\mathcal{Z}}_{g,v}^t\}$ will result in a null matrix, while for other variations $v'$, the subtraction results in matrices containing elements with $1$, $0$ and $-1$. By employing this feature, constraint (\ref{PF1}) can be linearized as 
\begin{eqnarray}\label{LNC1}
	\textstyle\sum_{j'\in\mathds{C}_j}\boldsymbol{P}_{jj'k}^{t}=\boldsymbol{P}_{ijk}^{t}-\boldsymbol{p}_{j}^{t}+\boldsymbol{P}_{j}^{t,\text{RES}}-\boldsymbol{\mathcal{Y}}_{ij}^t+\dot{\boldsymbol{p}}_{ijk}^{t},
\end{eqnarray}
\begin{eqnarray}\label{LNC2}
	\boldsymbol{\mathcal{Y}}_{ij}^t\leq \zeta_{g,v}^t\boldsymbol{1}M+\boldsymbol{r}_{ij,v}^{t}\tilde{\boldsymbol{I}}_{ijk}^{t},\forall v\in\mathds{V}, (i,j)\in\mathds{L}_g,
\end{eqnarray}
\begin{eqnarray}\label{LNC3}
	\boldsymbol{\mathcal{Y}}_{ij}^t\geq-\zeta_{g,v}^t\boldsymbol{1}M+\boldsymbol{r}_{ij,v}^t\tilde{\boldsymbol{I}}_{ijk}^{t},\forall v\in\mathds{V}, (i,j)\in\mathds{L}_g,
\end{eqnarray}
\begin{eqnarray}\label{LNC4}
	\zeta_{g,v}^t\geq \{\boldsymbol{\mathcal{S}}_g^t-{\boldsymbol{\mathcal{Z}}}_{g,v}^t\}_{\varphi\phi}, \forall v\in\mathds{V}, \varphi,\phi\in \Phi,
\end{eqnarray}
\begin{eqnarray}\label{LNC5}
	\textstyle\sum_{v\in\mathds{V}}\zeta_{g,v}^t\leq 5.
\end{eqnarray}
where $\zeta_{g,v}^t$ is an auxiliary variable introduced for the reordering selection. When $\zeta_{g,v}^t=0$, it means variation $v$ is selected at time $t$, and all the lines downstream switchgear $g$ must be reordered following ${\boldsymbol{\mathcal{Z}}}_{g,v}^t$. Constraints (\ref{LNC2})-(\ref{LNC3}) ensure that the auxiliary variable  $\boldsymbol{\mathcal{Y}}_{ij}^t=\boldsymbol{r}_{ij,v}^{t}\tilde{\boldsymbol{I}}_{ijk}^{t}$, while relax the constraints for all other variations $v'$. Constraints  (\ref{LNC4})-(\ref{LNC5}) implies that only one variation is selected at one time $t$. Note that the same linearization technique is applied for constraints (\ref{PF2})-(\ref{PF3}).

\section{Case Study}

In this section, the test system for case studies is described. The simulation of the proposed underground PDS restoration strategy using IBRs is conducted. The results of inrush current, ferroresonance, and phase swapping regarding  challenges associated with underground PDSs are presented and discussed. Comparative cases are analyzed against baseline strategies.

\subsection{Test System Setup}

For case studies, the underground PDS is established based on IEEE 123-Node Test Feeder, operating at a nominal voltage of $4.16$kV. A detailed illustration is shown in Fig. \ref{IEEE123}. The main feeder is composed of three-phase overhead lines, represented by orange solid lines. And, the laterals are single, two, or  three-phase underground cables, represented by triple-lines colored orange, blue, black, respectively, with a shunt capacitance of 178nF/mile. The underground PDS contains 14 pad-mounted smart switchgears, each associated with a downstream underground lateral area. Also, it includes 5 switches installed along the main feeder for sectionalization and tie-line operations. In addition, a total of 3 ESSs, 3 PVs and 4 WTs are integrated in the network, interfaced with the grid via three-phase inverters. The ESSs are grid-forming with an energy capacity of 3,000 kWh, 5,000 kWh and 8,000 kWh, respectively. Both the PVs and WTs are grid-following, with their hourly forecasted active output derived from \cite{abeykoon2018effect,de2012optimal}. The peak load demand is scaled to 1.0 p.u. as the same as the demand of IEEE 123-Node Test Feeder. Each node with load is supplied through a pad-mounted distribution transformer. The restoration begins at 7:00 AM, and is considered over 24 hours without power support from the bulk system.

\vspace{-10pt}

\subsection{Results Discussion and Comparison}

\begin{figure}[t]
	\centering
	\includegraphics[width=3.4in]{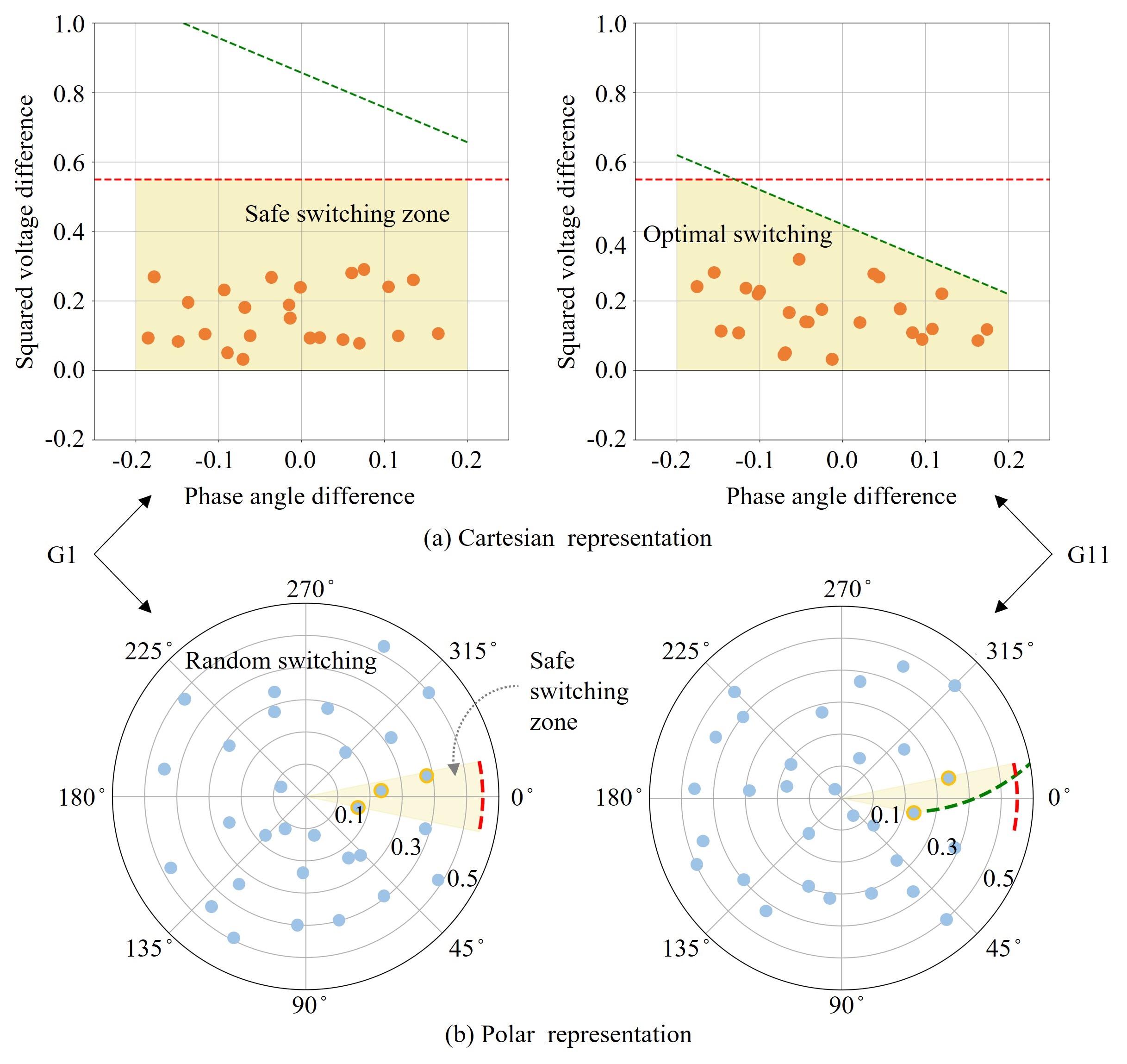}
	\caption{Results of inrush current when closing switchgears.}
	\label{IRC_Analysis}\vspace{-10pt}
\end{figure}

\subsubsection{Overall Performance Analysis}

The distribution network after topology reconfiguration is presented in Fig. \ref{Recfig}(a). Specifically, a total of three microgrids are formed with ESSs serving as the roots. Normally opened switches $(54,94)$ and $(151,300)$ are operated to enable load transfer. Microgrids from \#1 to \#3 comprise five, four, and five smart pad-mounted switchgears, respectively, with corresponding underground cable lengths of 1.02 miles, 1.11 miles, and 1.41 miles. The results indicate that the total weighted restored load over 24-hours is $30,221$kWh. During the restoration, the total renewable generation reaches 28,191 kWh, including 78.3\% of solar power and 21.7\% of wind power. The total renewable energy transition considering ESS charging and discharging is $21,153$kWh. Fig. \ref{Recfig}(b) depicts the ESS SOC, charging and discharging profiles corresponding to each microgrid. It can be observed that the ESSs primarily charge during periods of surplus renewable generation, particulaly between 9:00 AM and 12:00 AM. In the evening, after 17:00 PM, the ESSs tend to discharge to support peak demand. After midnight, when renewable generation declines, the load is predominantly supplied by the energy stored in the ESSs.

\subsubsection{Inrush Current Analysis} To demonstrate the control of inrush current, the representative results of switchgears G1 and G11 are presented in Fig. \ref{IRC_Analysis}(a). Specifically, the shaded regions enclosed by the dotted lines represent the allowable switching zones, within which a safe energization of underground cables is guaranteed. The blue line denotes the upper bound of inrush current based on the limitation $\mathcal{I}_{ij}^\text{max}$. The red line denotes the upper bound on the squared voltage difference $\Delta V_{ij}$. The left and right boundaries of the region are determined by a small phase angle difference of $\pm 10^{\circ}$. In addition, the orange dots represent the switching decisions at each time obtained by the optimization, which will be sent to the smart switchgears. We can observe that all the decisions lie within the shaded region, confirming that all the inrush currents are  under control. For example, the worst inrush currents of G1 and G11 are 1.59 p.u. and 1.83 p.u., respectively, each well below the threshold of 2.0 p.u. Moreover, in Fig. \ref{IRC_Analysis}(a), we can see that the region is dominated by the red line for switchgear G1, whereas by the green line for switchgear G11. 
This difference arises because the total length of underground laterals downstream of G1 is 0.42 miles, while for G11 it is 0.88 miles. In other words, the length of underground networks connected with G11 is much longer than that of G1, resulting in a larger capacitance with a higher risk of severe inrush during energization. 

To further highlight the importance of inrush current constraints, a total of thirty random points of squared voltage and phase angle differences are generated. The comparative results are plotted in Fig. \ref{IRC_Analysis}(b)  using polar coordinates. We can observe that a substantial number of switchgear operation fall outside the switching zone, indicating a higher likelihood of violating the maximum inrush current limits. Another observation from Fig. \ref{IRC_Analysis}(b) is that the region for safe operation is  small, thereby inrush current control is necessary.

\begin{figure}[t]
	\centering
	\includegraphics[width=3.4in]{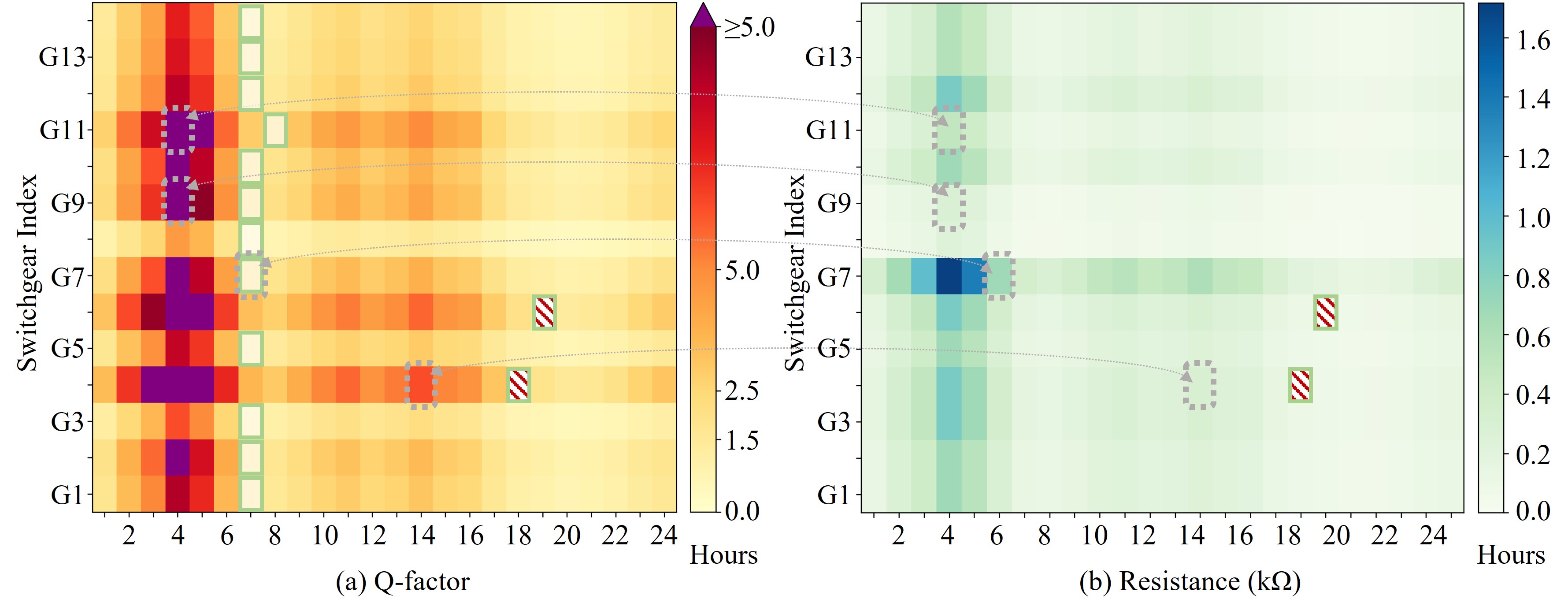}
	\caption{Results of ferroresonance when energizing transformers.}
	\label{FRR}\vspace{-10pt}
\end{figure}

\subsubsection{Ferroresonance Results Analysis}

The simulation results regarding ferroresonance are depicted in Fig. \ref{FRR}. Specifically, in Fig. \ref{FRR}(a), the values of Q-factor for switchgears are presented over 24-hours. A lower value in light yellow indicates a safer switchgear operation, whereas a higher $\mathcal{Q}$ that tends to be red signifies a higher likelihood of ferroresonance that can damage distribution transformers. Assuming that the restoration begins at 7:00 AM, then the green boxes is the first time when the Q-factors for each switchgear satisfy the safe operating condition. It is worth emphasizing that the operation for G11 cannot be performed immediately after restoration starts. Instead, it must be delayed until 8:00 AM to ensure a safety operation. While, such delays are constrained to prevent excessive waiting times. For example, the first operating times for G4 and G6 happen at 18:00 PM and 19:00 PM, both exceeding the delay threshold of 2 hours. Then, crews are dispatched to use portable protective air gaps to avoid waiting for prolonged natural damping. Also, the damping resistance downstream switchgears are presented in Fig. \ref{FRR}(b). By contrast with Fig. \ref{FRR}(a), we can observe that a lower resistance corresponds to a lower $\mathcal{Q}$ generally. Yet, the Q-factor is also significantly influenced by the capacitance of underground cables. For example, despite the strong damping, switchgear G11 still exhibits a higher Q-factor because of the long length and high capacitance. The results validate that the proposed ferroresonance model, and highlights the importance of considering damping levels. That is the damping resistance offers a conductive path through which oscillatory energy can be dissipated as heat, mitigating the ferroresonance conditions. For comparison, a simulation is further conducted for the case without considering ferrorensoance. While it results in a higher restored load of $30,441$ kWh, the outcome is unreliable, as it overlooks the risk of damaging distribution transformers due to uncontrolled switching actions.

\subsubsection{Phase Swapping Results Analysis}

The results related to phase swapping are listed in Table \ref{Swapping_results}, which demonstrates the effectiveness of phase swapping in mitigating load imbalance. Specifically, prior to restoration, considerable phase unbalance can arise when following the original phase arrangement. For example, microgrid \#1 exhibits the highest deviation, with a load distribution of 42.5\% on phase $a$, 17.7\% on phase $b$ and 39.8\% on phase $c$. This is because a disproportionate number of loads are connected to single phases $a$ and $c$. By utilizing smart switchgears for phase swapping, a more balanced load allocation can be achieved. For example, microgrid \#1 obtains an improved distribution of 31.4\% on phase $a$, 33.0\% on phase $b$, and 35.6\% on phase $c$. And, the deviation of phase loading is reduced from 0.111 to 0.017. In addition, the results of microgrid \#3 imply that phase swapping will not be conducted if the original load arrangement already satisfies the per-phase power constraints imposed by ESSs. For comparison, a simulation is conducted with phase swapping disabled. The results show that the multi-period restored load is 30,221kWh when phase swapping is allowed, while it decreases to 29,752kWh without it. This reduction is mainly due to more frequent load curtailments caused by violations of the ESS’s per-phase active power limits when phase imbalance is not addressed.

\section{Conclusion}

In this paper, a comprehensive restoration framework is proposed for underground PDSs leveraging IBRs. Three models are developed to address the unique challenges of underground PDSs, including an underground cable energization model that quantifies inrush currents, a distribution transformer energization model that considers ferroresonance risks, and a phase-swapping model that addresses load imbalance. Moreover, the models are integrated in an MINLP problem with the objective of maximizing the total multi-period restored load. A permutation-based linearization technique is  proposed. Case studies on IEEE 123-Node Feeder validate that the proposed strategy can limit inrush currents below maximum thresholds, enable transformer energization only when ferroresonance is acceptable, and reduce load imbalance compared to baselines.

Future research directions can extend the proposed framework to include frequency dynamics during underground PDS restoration. A swing-equation model can be integrated to quantify frequency nadir and rate-of-change-of-frequency (RoCoF) following underground cable vs overhead line energization. To mitigate frequency deviations, a closed-form control strategy requires further investigation, with coordinated support from grid-forming and grid-following inverters.

\begin{table}[!t]
	\renewcommand{\arraystretch}{1.4}
	\setlength{\tabcolsep}{7.5pt}
	\caption{Results of Phase Swapping Performance}
	\label{Swapping_results}
	\centering
	\begin{tabular}{c | c |c c c| c }
		\hline\hline
		{Index} & {Type} & {a} & {b} & {c} & {Deviation}\\
		\hline\hline
		\multirow{2}{*}{MG\#1} & Original & 42.5\% & 17.7\% & 39.8\% & 0.111  \\
		& Swapping & 31.4\% & 33.0\% & 35.6\% & 0.017 \\
		\hline
		\multirow{2}{*}{MG\#2} & Original & 42.3\% & 31.7\% & 26.0\% & 0.068  \\
		& Swapping & 35.60\% & 31.2\% & 33.20\% & 0.018  \\
		\hline
		\multirow{2}{*}{MG\#3} & Original & 28.9\% & 35.6\% & 35.5\% & 0.031  \\
		& Swapping & 28.9\% & 35.6\% & 35.5\% & 0.031  \\
		\hline\hline
	\end{tabular}
\end{table}

\bibliographystyle{IEEEtran}\bibliography{ref}

\end{document}